\documentclass[pra,aps,twocolumn]{revtex4-1}

\usepackage{graphicx}
\usepackage{amsmath}
\begin{document}
\title{Holographic Code Rate}
\author{Noah Bray-Ali}
\email{nbrayali1@csudh.edu}
\affiliation{Department of Physics, California State University, Dominguez Hills, California 90747 USA}
\author{David Chester}
\affiliation{Department of Physics and Astronomy, University of California, Los Angeles, California 90095 USA}
\affiliation{Quantum Gravity Research, Los Angeles, California 90290 USA}
\author{Dugan Hammock}
\affiliation{Quantum Gravity Research, Los Angeles, California 90290 USA}   
\author{Marcelo M. Amaral}
\affiliation{Quantum Gravity Research, Los Angeles, California 90290 USA}   
\author{Klee Irwin}
\affiliation{Quantum Gravity Research, Los Angeles, California 90290 USA}   
\author{Michael F. Rios}
\affiliation{Dyonica, ICMQG, Los Angeles, California 90032 USA}
\date{\today}
\begin{abstract}
Holographic codes grown with perfect tensors on regular hyperbolic tessellations using an inflation rule protect quantum information stored in the bulk from errors on the boundary provided the code rate is less than one.  Hyperbolic geometry bounds the holographic code rate and guarantees quantum error correction for codes grown with any inflation rule on all regular hyperbolic tessellations in a class whose size grows exponentially with the rank of the perfect tensors for rank five and higher.  For the tile completion inflation rule, holographic triangle codes have code rate more than one but all others perform quantum error correction.
\end{abstract}
\maketitle
\section{\label{intro}Introduction}
%%6 pages, 3 tables, 4 figures

Holographic quantum error-correcting codes\cite{pastawski2015,jahn2019} merge quasi-crystals\cite{boyle2018} and hyperbolic
geometry\cite{thurston1997,kollar2019} with quantum information\cite{shor1996,calderbank1996,steane1996,aharonov1997,cleve1999,reed2012,helwig2012,helwig2013,harris2018} and holography\cite{brown1986,witten1988,strominger1998,witten1998,vidal2008,qi2013,yoshida2013,latorre2015,pastawski2017}.
One places rank-$(p+1)$ perfect tensors $T_{a_1a_2\ldots a_{p}a_{p+1}}$
on the $p$-sided tiles of a tessellation of the hyperbolic plane
and contracts the tensors along the edges where tiles meet: this
leaves a single ``bulk'' index uncontracted for each
tile\cite{pastawski2015}.  Starting from some simply connected set of
seed tiles, we grow the holographic code, layer by layer, using an
inflation rule\cite{boyle2018}.

The physical degrees of freedom of the code live on the boundary of the growing
tile set on the quasi-crystal formed by the dangling edges of the
tiles of the last layer\cite{pastawski2015}.  The logical degrees of freedom of the code live in the bulk of the tile set on the tiles themselves.  The perfect tensors map
the physical Hilbert space isometrically to the logical
Hilbert space\footnote{To see the isometry, split the perfect tensor index list into two pieces $A=\{a_{i_1},a_{i_2},\ldots a_{i_k}\}$ and $B=\{a_{i_{k+1}},a_{i_{k+2}},\ldots a_{i_{p+1}}\}$, with $k=|A|$ less than or equal to half of $p+1$ the rank of the perfect tensor.  Then view the perfect tensor as a linear map $T_{AB}$ from the Hilbert space $H_A$ to the Hilbert space $H_B$.  For any such split, the perfect tensor preserves the overlap between states up to a proportionality constant: $\sum_B (T_{AB})^*T_{A'B}=1/{\rm dim}(H_A)\delta_{AA'}$, where, ${\rm dim}(H_A)$ is the dimension of $H_A$.  For example, when the dimension of $H_A$ and $H_B$ match, then the linear map $T_{AB}$ is proportional to a unitary operator.  For the purposes of holography and quantum error-correction, the tensors need only preserve the overlap between states for index splits which have the ``block'' form $i_j=i_{j+1}+1$ modulo $(p+1)$ for $j=1,2,\ldots,p+1$.  In fact the holographic code rate for a given hyperbolic tessellation is the same for codes grown with perfect and block perfect tensors\cite{harris2018}.}.

\begin{figure}[ht]
\includegraphics[width=3in,height=2in]{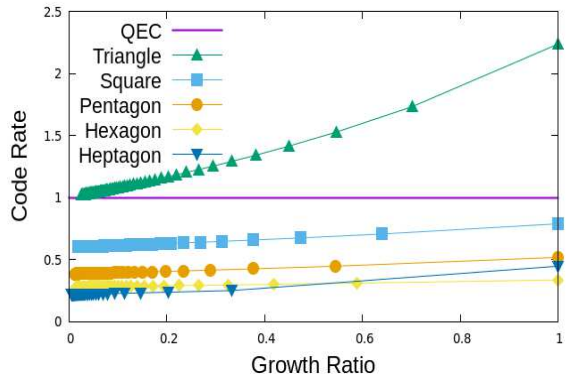}

  \caption{Quantum error correction threshold (QEC) together with code rates of holographic codes grown with tile completion on the $\{p,q\}$-tiling of the hyperbolic plane by regular $p$-gons meeting $q$ around a vertex $(1/p+1/q<1/2)$ plotted as a function of the ratio of the growth rate bound set by hyperbolic geometry to the growth rate for $p=3,4,5,6,7$ (Square, Pentagon, Hexagon, Heptagon).}
\label{tilecoderate}
\end{figure}

\begin{figure}[ht]
\includegraphics[width=3in,height=2in]{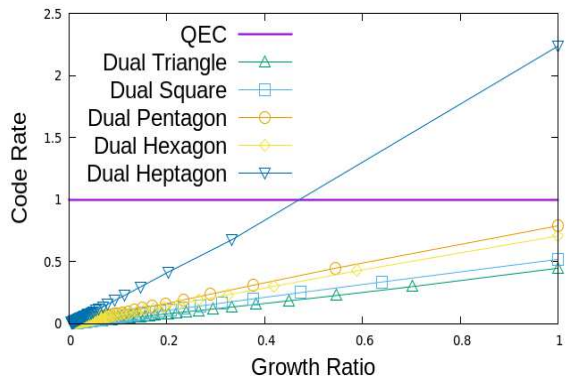}
  \caption{Quantum error correction threshold (QEC) together with code rates of holographic codes grown with tile completion on the $\{p,q\}$-tiling of the hyperbolic plane by regular $p$-gons meeting $q$ around a vertex $(1/p+1/q<1/2)$ plotted as a function of the ratio of the growth rate bound set by hyperbolic geometry to the growth rate for $q=3,4,5,6,7$ (Dual Triangle, Dual Square, Dual Pentagon, Dual Hexagon, Dual Heptagon).}
\label{dualtilecoderate}
\end{figure}

The quantum error-correcting property of a holographic code follows from a remarkable fact about hyperbolic geometry\footnote{The quantum error correction property we have in mind is simply the existence of a finite, non-zero threshold for the fraction of quantum erasures of physical degrees of freedom below which the logical degrees of freedom can be recovered, in the limit of large code size.  A necessary condition for this quantum error-correcting property is that the code rate be less than one. Conversely, numerical simulations suggest the existence of the non-zero erasure threshold for holographic codes with code rate less than one\cite{pastawski2015,harris2018}.}.  For a given growth rule $\tau(p,q)$ on the $\{p,q\}$-hyperbolic
tessellation with regular $p$-sided tiles meeting $q$ around a vertex,
there exists a finite code rate
\begin{equation}
 \chi_{\tau(p,q)}=\lim_{n\rightarrow\infty}\left(\frac{N_{bulk}}{N_{boundary}}\right),
\label{coderatedef}
\end{equation}
where, $N_{bulk}$ is the number of logical degrees of freedom, $N_{boundary}$ is
the number of physical degrees of freedom, and $n$ is the number of layers of the
code.  In particular, there exist growth rules and tilings such that
the code rate $\chi_{\tau(p,q)}$ is less than one (Table \ref{chitable},Fig.~\ref{tilecoderate},~\ref{dualtilecoderate}).  For such
holographic codes, quantum erasures of a non-zero fraction of
physical degrees of freedom on the boundary do not harm the quantum information
stored in the logical degrees of freedom of the bulk\cite{pastawski2015,harris2018}.
\begin{table}[tbp]
\begin{ruledtabular}
\begin{tabular}{lllll}
$p$ & $q$& $\chi_{\tau C(p,q)}$ & $\chi_{p,q}$ & $\chi_{\tau C(p,q)}/\chi_{p,q}$\\
%\colrule
3 & 7 & 2.236 & 2.430 & 0.920 \\
4 & 5 & 0.789 & 0.998 & 0.790 \\
5 & 4 & 0.519 & 0.676 & 0.768 \\
7 & 3 & 0.447 & 0.541 & 0.826\\
\end{tabular}
\end{ruledtabular}
\caption{Code rate $\chi_{\tau(p,q)}$ of holographic codes grown with tile completion on the $\{p,q\}$-tiling of the hyperbolic plane by regular $p$-gons meeting $q$ around a vertex together with $\chi_{p,q}$ code rate bound from hyperbolic geometry.  The code rate is greater than one for triangle codes, is less than than one for $p$-gon codes with $p$ greater than three, and obeys the bound for all $p$.}
\label{chitable}
\end{table}

We have found a simple geometric upper bound $\chi_{p,q}$
on the code rate for all $p$ and $q$ such that
$\frac{1}{p}+\frac{1}{q}<\frac{1}{2}$, which is the condition that the tiling lives
in the hyperbolic plane.  The result takes the following form:
\begin{equation}
  \chi_{p,q}=\frac{\ell_{p,q}}{a_{p,q}},
  \label{bound}
\end{equation}
where, $\ell_{p,q}=2\cosh^{-1}(\cos(\pi/p)/\sin(\pi/q))$ is the length of the side of the regular $p$-gon tile and $a_{p,q}=2\pi p(1/2-1/p-1/q)$ is its area\cite{thurston1997}.  Notice that the condition $\frac{1}{p}+\frac{1}{q}<\frac{1}{2}$ is necessary and sufficient to make the area of the tiles $a_{p,q}$ positive and the length of their sides $\ell_{p,q}$ real.  Remarkably, for all $p>3$, there exists an exponentially growing range $(q_{min}(p),q_{max}(p))$ such that, for $q$ in this range, the simple geometric bound $\chi_{p,q}$ guarantees quantum error correction for all holographic codes grown with any growth rule from any simply connected seed tiles in the hyperbolic plane.    

Sec.~\ref{sec:coderatebound} finds the code rate bound using hyperbolic geometry.  In Sec.~\ref{sec:bestcodebound}, we find the best code rate bound and the range of codes for which the bound guarantees quantum error correction.  Sec.~\ref{sec:coderate} shows the holographic code rate for the tile completion growth rule obeys the bound from hyperbolic geometry (Fig.~\ref{coderatio},\ref{dualcoderatio}).  Using this growth rule, we find that all codes with $p>3$ perform quantum error correction but the code rate is greater than one for holographic triangle codes (Fig.~\ref{tilecoderate},\ref{dualtilecoderate}).  In Sec.~\ref{sec:tilecompletion} we provide details of the quasi-crystalline interpretation which allows the holographic code rate to be computed for all regular tilings using the tile completion growth rule and we make contact with other growth rules and interpretations introduced in the literature\cite{pastawski2015,boyle2018}.%\\

%\bigskip
\section{\label{sec:coderatebound}Code Rate Bound}

%%%
The code rate of holographic codes has an upper bound from hyperbolic geometry.  To show this, we use the following result from plane geometry, known as the isoperimetric inequality\cite{courant1953}:
\begin{equation}
A(4\pi+kA)\le L^2,
\label{isoperimetric}
\end{equation}
where, $L$ is the length of the curve bounding a region of the plane with area $A$ and $k=1,0,-1$ for the hyperbolic, Euclidean, and elliptical plane, respectively.  The name of the inequality refers to the isoperimetric problem of finding the curve with given length which bounds the largest area in the plane.  The classical solution is simply the circle, consisting of all the points with fixed geodesic distance from a given point in the plane. Indeed, the circle saturates Eq.~(\ref{isoperimetric}), as one can show using plane geometry.\footnote{The circle of radius $s$ in the hyperbolic plane has circumference $L(s)=2\pi\sinh(s)=4\pi\sinh(s/2)\cosh(s/2)$ and area $A(s)=4\pi\sinh^2(s/2).$  Using the relation between hyperbolic sine and cosine $1+\sinh^2(s/2)=\cosh^2(s/2)$, we find the circumference and area of the circle in the hyperbolic plane obey the equality $L(s)^2=A(s)(4\pi+A(s))$, saturating the bound in Eq.~(\ref{isoperimetric}).}  

Let us apply this to the holographic code growing on the regular $\{p,q\}$-tiling of the hyperbolic plane.  At a given layer, the code spans $N_{bulk}$ tiles in the bulk with $N_{boundary}$ edges along its boundary.  The length of the boundary $L=\ell_{p,q}N_{boundary}$ and the area of the bulk $A=a_{p,q}N_{bulk}$ follow from the fact that the tiles are regular with each edge having the same length $\ell_{p,q}$ and each tile having the same area $a_{p,q}$.  

Using the isoperimetric inequality, we find the code rate bound for holographic codes.  First, plug the length $L=\ell_{p,q}N_{boundary}$ and area $A=a_{p,q}N_{bulk}$ into the isoperimetric inequality for the hyperbolic plane.  Next, take the square root of both sides.  At last, divide both sides by $N_{boundary}a_{p,q}$ to find:
\begin{equation}
\frac{N_{bulk}}{N_{boundary}}\sqrt{1+\frac{4\pi}{N_{bulk}a_{p,q}}}\le \frac{\ell_{p,q}}{a_{p,q}}.  
\label{prelimit}
\end{equation}
Finally, to establish the code rate bound (\ref{bound}), we pass to the limit of infinite layer.  The number of tiles $N_{bulk}$ goes to infinity in this limit and the left hand side of the inequality (\ref{prelimit}) becomes the code rate, establishing the bound on code rate from hyperbolic geometry. 

\section{\label{sec:bestcodebound}Best Code Rate Bound}

The results for the best code rate bound $\chi_{p,q_{opt}(p)}$ and optimal number of tiles $q_{opt}(p)$ for $3\le p\le 10$ are shown in Table \ref{bestbound}.  For $8\le p\le 11$, the
optimum bound comes with five tiles around a vertex.  For tiles with
more sides $12\leq p\leq 30$ the optimum number of tiles drops to four
while the best code rate bound falls from $\chi_{12,4}=0.132$ to
$\chi_{30,4}=0.043$.  For tiles with a huge number of sides $p>30$ the
optimum bound occurs for three tiles meeting around a vertex, the
least possible, and the optimum bound $\chi_{p,3}$ falls to zero
$O(1/p)$ inversely with $p$ as $p$ goes to infinity.
\begin{table}[tbp]
\begin{ruledtabular}
\begin{tabular}{lllllllll}
%\colrule 
$p$ & 3 & 4 & 5 & 6 & 7 & 8 & 9 & 10\\   
 $q_{opt}(p)$ & 14 & 9 & 7 & 6 & 6 & 5 & 5 & 5  \\
 $\chi_{p,q_{opt}(p)}$ & 1.614 & 0.776 & 0.500 & 0.365 & 0.285 & 0.233 & 0.196 & 0.169  
 \end{tabular}
\end{ruledtabular}
\caption{Best code rate bounds $\chi_{p,q_{opt}(p)}$ and number of
   tiles $q_{opt}(p)$ meeting around a vertex for holographic codes
   grown with regular $p$-sided tiles on the hyperbolic plane.}
\label{bestbound}
\end{table}

We note that the best code rate bound is {\em less} than one for all
$p$ greater than three.  This shows that {\em any} perfect tensor of rank five and
higher has at least one hyperbolic tessellation on which {\em every} holographic
code grown with the tensor has code rate less than one.  Thus we are
guaranteed by hyperbolic geometry the existence of holographic codes
that perform quantum error-correction, provided we can construct
perfect tensors with rank five and higher.

Further, there exists the range
$(q_{min}(p),q_{max}(p))$ within which $\chi_{p,q}<1$ for all $p>3$.  In particular, we report this range
together with the code rate bounds $\chi_{p,q}$ and an analytic
estimate $q_{1}(p)$ for $q_{max}(p)$ for $4\le p\le 7$ in Table
\ref{boundrange}.  Notice that $\chi_{p,3}$ decreases with $p$ and,
since the bound $\chi_{7,3}=0.541$ is less than one, it follows that
the minimum $q_{min}(p)$ must be three for all $p\ge 7$.  The maximum
$q_{max}(p)$ clearly grows quickly already for the small $p$ in Table
\ref{boundrange}.  Meanwhile the minimum $q_{min}(p)=1+\lfloor 2+4/(p-2)\rfloor$ is simply the minimum number of tiles around a vertex allowed by hyperbolic geometry: the range includes all geometrically allowed tilings at the lower end.

\begin{table}[tbp]
\begin{ruledtabular} 
\begin{tabular}{lllll}
%\colrule
$p$ & 4 & 5 & 6 & 7 \\   
 $q_{min}(p)$ & 5 & 4 & 4 & 3  \\
% $\chi_{p,q_{min}(p)}$ & 0.998 &  &  &  \\ 
$q_{max}(p)$ & 36 & 199 & 952 & 4468 \\
%$\chi_{p,q_{max}(p)}$ & 0.997 & & &
$q_{1}(p)$ & 45 & 216 & 971 & 4491 
\end{tabular}
\end{ruledtabular}
\caption{Boundaries of the range ($q_{min}(p),q_{max}(p))$ within
  which the code rate bound $\chi_{p,q}$ guarantees quantum error
  correction for {\em all} holographic codes, and the asymptotically
  exact estimate $q_1(p)=\pi\cosh(\pi(p-2)/2)/\cosh(\pi/p)$ for
  $q_{max}(p)$.}
\label{boundrange}
\end{table}

For large $p$, we find how $q_{max}(p)$ grows by setting
$\chi_{p,q}=1$ and expanding
$\sin(\pi/q)=\pi/q+\ldots$ and $a_{p,q}=\pi (p-2)/2+\ldots$, where,
the omitted terms are sub-leading in $1/q$.  The result is the
asymptoticaly exact estimate $q_{1}(p)=\pi \cosh(\pi
(p-2)/2)/\cos(\pi/p)$ which grows $O(\exp(\pi p/2))$ exponentially for
large $p$.  We see in Table 2 that already for small $p$ the estimate
gives a good approximation to $q_{max}(p)$.

\section{\label{sec:coderate}Holographic Code Rate}
\begin{figure}
\includegraphics[width=3in,height=2in]{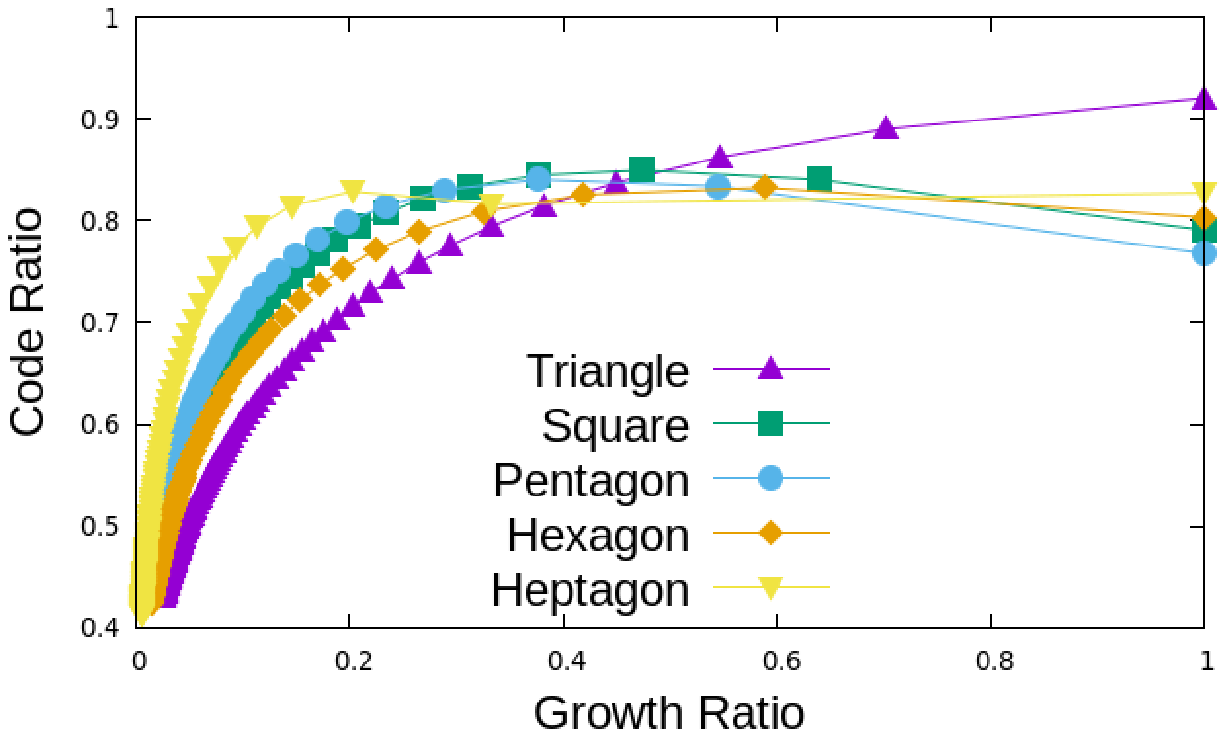}
  \caption{Ratio of the code rate to the code rate bound from hyperbolic geometry for holographic codes grown with tile completion on the ${p,q}$-tiling of the hyperbolic plane by regular $p$-gons meeting $q$ around a vertex $(1/p+1/q<1/2)$ plotted as a function of the ratio of the growth rate bound from hypebolic geometric to the growth rate for $p=3,4,5,6,7$ (Square, Pentagon, Hexagon, Heptagon).}
%\label{tilecodebound}
\label{coderatio}
\end{figure}

\begin{figure}[ht]
  \includegraphics[width=3in,height=2in]{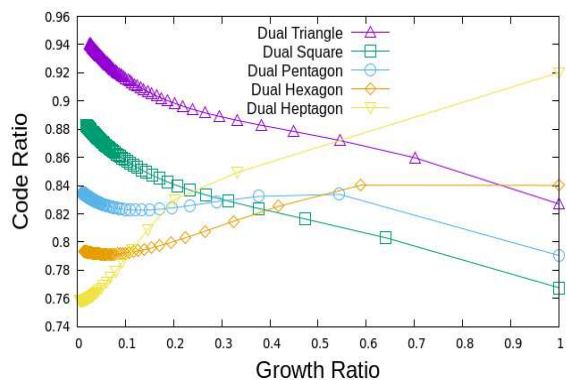}
  \caption{Ratio of the code rate to the code rate bound from hyperbolic geometry for holographic codes grown with tile completion on the ${p,q}$-tiling of the hyperbolic plane by regular $p$-gons meeting $q$ around a vertex $(1/p+1/q<1/2)$ plotted as a function of the ratio of the growth rate bound from hyperbolic geometric to the growth rate for $q=3,4,5,6,7$ (Dual Triangle, Dual Square, Dual Pentagon, Dual Hexagon, Dual Heptagon).}
\label{dualcoderatio}
\end{figure}

The code rate of the holographic code grown on {\em any} hyperbolic
tessellation with {\em any} growth rule obeys the upper bound from
hyperbolic geometry.  We check this fact for the tile completion growth rule for which the code rate may be computed analytically for {\em all} regular tilings of the hyperbolic plane (Table~\ref{chitable} and Fig.~\ref{coderatio},\ref{dualcoderatio}).  Along the way, we show that holographic triangle codes have code rate greater than one and holographic $p$-gon codes with $p$ greater than three have code rate less than one (Fig.~\ref{tilecoderate},\ref{dualtilecoderate}).

Here is how to grow the holographic code using the tile completion growth rule\cite{boyle2018}.  Start from a simply connected set of seed tiles
which form the zero-th layer.  The first layer of the code is made of
all the tiles that share a vertex with a seed tile.  Similarly, the second layer consists of all the tiles which share a vertex with a tile in the first layer but which are not in the seed layer, and so on, layer by layer.\footnote{Tile completion on the $\{p,q\}$-tiling gives the same code as vertex completion on the $\{q,p\}$-tiling by duality.  The duality transformation joins the centers of the regular $p$-gon tiles with segments.  These segments form faces with $q$ segments around a face, and $p$ faces meeting at each point where the segments end.}

The basic fact about the code rate $\chi_{\tau C(p,q)}$ for holographic codes grown with tile completion on the $\{p,q\}$-tiling of the hyperbolic plane is that it decreases as a function of $p$ at fixed $q$ and as a function of $q$ at fixed $p$ (Section~\ref{sec:tilecompletion}).  Here we will simply use this fact to show that the code rate obeys the bound from hyperbolic geometry and that all codes with $p>3$ perform quantum error correction.

Given that the code rate $\chi_{\tau C(p,q)}$ falls with $q$ at fixed $p$ and with $p$ at fixed $q$, it is enough to show the code rates are less than one for just three codes (Table \ref{chitable}): the heptagon code grown on the $\{7,3\}$-tiling, the pentagon code grown on the $\{5,4\}$-tiling, and the square code grown on the $\{4,5\}$-tiling.  These tilings have code rate less than one and we turn now to showing that they dominate the code rate of holographic $p$-gon codes with $p$ greater than three.  

To show that all holographic codes grown with tile completion have code rate bounded by one of these three codes we begin by analyzing codes with seven or more edges around a tile.  Codes with more than seven edges around a tile have
lower code rate than the heptagon code with the same number of tiles
meeting around a vertex, since the code rate falls when we increase
the number of sides around a tile while keeping fixed the number of
tiles around a vertex.  Similarly, these heptagon codes have lower
code rate than the heptagon code grown with three tiles around a
vertex, since the code rate falls with increasing number of tiles
around a vertex while keeping fixed the number of edges around a tile.  The code rate $\chi_{\tau C(7,3)}=0.447$ is less than one, and so we have shown that the holographic $p$-gon codes with $p$ greater than or equal to seven all have code rate less than one.

By the same reasoning, we analyze the code rate of codes with six edges around a tile and show that they are less than one. The hexagon code with four tiles around a vertex has the largest code rate among hexagon codes.  Now, this hexagon code has code rate which is lower than that of the pentagon code grown on the $\{5,4\}$-tiling since that tiling has the same number of tiles around vertex but a smaller number of edges around a tile.  So, for hexagon codes the code rate must be less than that of the pentagon code grown on the $\{5,4\}$ tiling of the hyperbolic plane.  

Finally, we consider the code rate of pentagon and square codes to show that they too have code rate less than one. The pentagon code
grown on the $\{5,q\}$-tiling with $q$ greater than four and the
square code grown on the $\{4,q\}$-tiling with q greater than five
have lower code rate than the pentagon code grown on the
$\{5,4\}$-tiling and the square code grown on the $\{4,5\}$-tiling,
respectively.  Again, the reasoning is that the code rate falls when we increase the number of tiles around a vertex for a fixed number of edges around a tile.  The code rates $\chi_{\tau C(4,5)}=0.789$ and $\chi_{\tau C(5,4)}=0.519$ are less than one, and thus we have shown that holographic square, pentagon, and hexagon codes have code rate less than one.  

Having shown that holographic codes grown with tile completion perform quantum error correction on all regular hyperbolic tessellations with tiles having more than three sides, we turn to checking the bound on the holographic code rate from hyperbolic geometry for all such codes including those grown on triangular tilings.  Recall that there exists a best code rate bound which occurs when we tune to the optimum number of tiles $q_{opt}(p)$ meeting around a vertex while keeping the number of sides $p$ around a tile fixed.  Now, the tile completion code rate falls with $q$ at fixed $p$, so the ratio of the code rate to the code rate bound has its maximum when the number of tiles $q_{max}(p)\le q_{opt}(p)$ is less than or equal to the number of tiles that gives the best code rate bound.  This is the key to the analysis as it presents us with a finite number of codes to analyze for each $p$.  

We break the analysis of the ratio of the code rate to the code rate bound further into two cases: codes with $p>30$ for which the best code rate bound occurs for the dual triangle code $\{p,3\}$ with three tiles around each vertex and codes with $p\le 30$ for which the best code rate bound comes from using a larger, but still finite number of tiles around each vertex.  For the $p>30$ case, we look at the ratio of the dual triangle code rate to the code rate bound from hyperbolic geometry.  It rises with $p$ and tends to the finite limit $\lim_{p\rightarrow\infty}(\chi_{\tau C(p,3)}/\chi_{p,3})=\pi/(3\ln 3)\approx 0.953$ which is less than one.  Thus the holographic $p$-gon codes with $p$ greater than thirty obey the code rate bound from hyperbolic geometry, as they must.  

The analysis of the ratio of the code rate to the code rate bound for codes with number of sides per tile $p\le 30$, requires us to search the finite number of codes with $q\le q_{opt}(p)$ to establish that the code rate bound is obeyed.  In fact, the ratio of code rate to code rate bound reaches its maximum among codes with $p\le 30$ for the holographic triangle code grown with tile completion on the $\{3,7\}$-tiling.  This code has the smallest growth rate among triangle codes and gives the ratio $\chi_{\tau C(3,7)}/\chi_{3,7}=2.236/2.430\approx0.920$, still less than one (Fig.~\ref{coderatio},\ref{dualcoderatio} and Table~\ref{chitable}).  Thus, in all cases the code rate of holographic codes grown with tile completion on regular tilings of the hyperbolic plane obeys the code rate bound from hyperbolic geometry.

\section{\label{sec:tilecompletion}Tile Completion Details}

We give here the details of the code rate of the holographic code grown with tile completion growth rule on the hyperbolic plane.  After a finite number of layers, the holographic code takes a quasi-crystal form with two unit cell types\cite{boyle2018}.  The cells each have one tile but differ in the number of dangling edges.  For codes grown with three tiles meeting around a vertex, the two types have $p-3$ and $p-4$ dangling edges while for codes grown with more than three tiles meeting around a vertex the two types of tiles have $p-3$ and $p-2$ dangling edges.  In either case, we find a growth rule relating the number of cells of each type in the current layer with the number of cells of each type in the next layer.  

The growth rule for holographic codes grown with tile completion on the $\{p,q\}$-tiling of the hyperbolic plane may be put in matrix form $M_{\tau C(p,q)}$\cite{pastawski2015,boyle2018}.  We compile the number of cells of each type into the growth vector $\vec{u}$ and find that the tile completion growth rule leads to the linear relationship $\vec{u}'=M_{\tau C(p,q)} \vec{u}$, where, $\vec{u}'$ is the vector containing the number of cells of each type after applying the growth rule.  Further, after a finite number of layers, the matrix becomes square with rank two, determinant $\det M_{\tau C(p,q)}$ equal to one, and integer matrix elements.  In sum, the growth matrix becomes part of the group $SL(2,Z)$ of rank two square matrices with unit determinant and integer coefficients\cite{boyle2018}.  

Explicitly, the growth matrix takes the following form for $p$ and $q$ both greater than three:
\begin{equation}
M_{\tau C(p>3,q>3)}=
\begin{pmatrix}
p-3 & (p-3)(q-3)-1\\
p-2 & (p-2)(q-3)-1
\end{pmatrix}
\label{growthrulegeneric}
\end{equation}
where, we express the growth matrix in the basis in which the tile vector is $\vec{t}=(1,1)^T$ and the edge vector $\vec{e}_{\tau C(p,q)}=(p-3,p-2)^T$.  These vectors list in a column the number of tiles and dangling edges, respectively, for each type of cell.  Similarly, for triangle codes we find the growth matrix for codes where the number of triangles meeting around a vertex $q$ is  greater than six:
\begin{equation}
M_{\tau C(3,q>6)}=
\begin{pmatrix}
0  & 1\\
-1 & q-4
\end{pmatrix}
\label{growthruletriangle}
\end{equation}
Here, the basis is such that the tile vector is $\vec{t}=(1,1)^T$ and the edge vector is $\vec{e}_{\tau C(3,q)}=(0,1)^T.$  Finally for dual triangle codes in which three tiles meet around each vertex, we find the growth matrix for tiles with the number of edges around each tile $p$ greater than six:
\begin{equation}
M_{\tau C(p>6,3)}=
\begin{pmatrix}
1  & p-6\\
1 & p-5
\end{pmatrix}
\label{growthruledualtriangle}
\end{equation}
Here, the basis is such that the tile vector is $\vec{t}=(1,1)^T$ and the edge vector is $\vec{e}_{\tau C(p,3)}=(p-4,p-3)^T.$  We notice that the determinant of these matrices is one while the trace is $2\gamma(p,q)=(p-2)(q-2)-2$ for all $p$ and $q$.

The growth rate of the holographic code grown with tile completion on the $\{p,q\}$-tiling of the hyperbolic plane is simply the largest eigenvalue of the growth matrix $M_{\tau C(p,q)}$\cite{boyle2018}:
\begin{equation}
\lambda_{\tau C(p,q)}=\gamma(p,q)+\sqrt{\gamma(p,q)^2-1},
\label{growthrate}
\end{equation}
where, $\gamma(p,q)=\frac{1}{2}{\rm tr} M_{\tau C(p,q)}=((p-2)(q-2)-2)/2$ is half the trace of the growth matrix.  Note that the growth rate is greater than one and is not rational for $1/p+1/q<1/2$.  Further, the growth rate grows with $p$ at fixed $q$ and with $q$ at fixed $p$ and, in fact, is symmetric under exchange of $p$ and $q$: $\lambda_{\tau C(p,q)}=\lambda_{\tau C(q,p)}$.  

Hyperbolic geometry provides a natural lower bound on the growth rate of holographic codes grown with tile completion.  The growth rate bound for triangle codes and dual triangle codes is $\lambda_{\tau C(3,7)}$, while for square codes and dual square codes it is $\lambda_{\tau C(4,5)}$. These bounds follow from the fact that the growth rate $\lambda_{\tau C(p,q)}$ is symmetric with respect to interchange of $p$ and $q$, grows with $p$ at fixed $q$ and with $q$ at fixed $p$, and that we must have $1/p+1/q<1/2$ for the tiling to live in the hyperbolic plane.  Similarly pentagon and dual pentagon codes have growth rate bound $\lambda_{\tau C(5,4)}$ and hexagon and dual hexagon codes have growth rate bound $\lambda_{\tau C(6,4)}$.  For $p\ge 7$, we find the $p$-gon code with three tiles around a vertex grows slowest giving growth rate bound $\lambda_{\tau C(p,3)}$.  Finally, for $q\ge 7$ the triangle code with $q$ tiles around a vertex gives the growth rate bound $\lambda_{\tau C(3,q)}$.

As the number of layers goes to infinity, the growth vector then tends to $\vec{u}_{\tau C(p,q)}$ the eigenvector of the growth matrix $M_{\tau C(p,q)}$ with eigenvalue equal to the growth rate\cite{pastawski2015}.  The number of cells at layer $n$ of each type forms a geometric sequence with ratio between consecutive terms equal to the growth rate and the initial value tending to the corresponding component of the growth vector $\vec{u}_{\tau C(p,q)}$.  Explicitly, we find the growth vector is $\vec{u}_{\tau C(p,3)}=(p-6,\lambda_{\tau C(p,3)}-1)$for codes where three tiles meet around each vertex.  For triangle codes, the growth vector becomes $\vec{u}_{\tau C(3,q)}=(1,\lambda_{\tau C(3,q)})$.  For codes with more than three tiles meeting around a vertex and more than three sides per tile, the growth vector takes the form $\vec{u}_{p,q}=((p-3)(q-3)-1, \lambda_{\tau C(p,q)}-(p-3))$.

Finally, the code rate of holographic codes grown with the tile rule on the
$\{p,q\}$-tiling of the hyperbolic plane by regular $p$-gons meeting
$q$ around a vertex, where, $p$ and $q$ are numbers such that
$1/p+1/q<1/2$ takes the form\cite{pastawski2015}:
\begin{equation}
\chi_{\tau C(p,q)}=\frac{\lambda_{\tau C(p,q)}}{\lambda_{\tau c(p,q)}-1}\frac{\vec{u}_{\tau C(p,q)}\cdot\vec{t}}{\vec{u}_{\tau C(p,q)}\cdot\vec{e}_{\tau C(p,q)}}.
\label{coderateform}
\end{equation}
Here, we have summed the geometric series $N_{bulk}=\sum_{k=1}^{n}\lambda_{\tau C(p,q)}^k (\vec{u}_{\tau C(p,q)}\cdot\vec{t})=\lambda_{\tau C(p,q)}^{n+1}/(\lambda_{\tau C(p,q)}-1)(\vec{u}_{\tau C(p,q)}\cdot\vec{t})+\ldots$ to obtain the number of tiles in the bulk at layer $n$, and omitted terms which are subleading in powers of the growth rate.  Similarly, the number of dangling edges on the boundary tends to $N_{boundary}=\lambda_{\tau C(p,q)}^n(\vec{u}_{\tau C(p,q)}\cdot\vec{e}_{\tau C(p,q)}$), as the number of layers $n$ goes to infinity.  Taking the ratio $N_{bulk}/N_{boundary}$ and passing to the limit of infinite layer number gives the code rate in Eq.~(\ref{coderateform}).  We note that this agrees with the form of the code rate found using a different growth rule in \cite{pastawski2015}.  In fact, the form applies for {\em any} growth rule which generates a quasi-crystal with a finite number of cell types after all but a finite number of layers\cite{boyle2018}. 

Using the explicit form for the code rate, we show that triangle codes have code rate greater than one.  We compute the triangle code rate in terms of the growth rate:
\begin{equation}
  \chi_{\tau C(3,q)}=\frac{\lambda_{\tau C(3,q)}+1}{\lambda_{\tau C(3,q)}-1}.
  \label{trianglecomp}
 \end{equation}
The fact that the growth rate is greater than one then gives the
result that the code rate for holographic triangle codes grown with
the tile rule must also be greater than one.  Along the way, we notice that the code rate $\chi_{\tau C(3,q)}$ of triangle codes drops as $q$ increases, since increasing $q$ increases the growth rate $\lambda_{\tau C(3,q)}$ which drives the code rate down to one, according to Eq.~(\ref{trianglecomp}).

Moving on to $p$-gon codes with $p$ greater than three, we show that the code rate $\chi_{\tau C(p,q)} $drops as $p$ increases at fixed $q$ and with increasing $q$ at fixed $p$.  We begin by noting that the code rate drops to zero as $1/p$ as $p$ goes to infinity at fixed
$q$.  To show this, we note the growth rate goes to infinity in this
limit and takes the factor $\lambda_{\tau C(p,q)}/(\lambda_{\tau C(p,q)}-1)$ to one.  Similarly, the edge vector becomes parallel to the tile vector
$\vec{e}_{\tau C(p,q)}=p\vec{t}$, up to corrections of order $1/p$.  Thus, the
growth vector cancels from the ratio
$(\vec{u}_{\tau C(p,q)}\cdot\vec{t}_{\tau C(p,q)})/(\vec{u}_{\tau C(p,q)}\cdot\vec{e}_{\tau C(p,q)})$
giving the result that the code rate fall to zero as $1/p$ as $p$ goes
to infinity at fixed $q.$

Continuting the analysis of the code rate for $p$-gon codes with $p$ greater than three, we note that the code rate falls to the non-zero, $q$-independent function
$(\vec{e}_{\tau C(p,q)}\cdot\vec{t})/(\vec{e}_{\tau C(p,q)}\cdot\vec{e}_{\tau C(p,q)})=((p-3)+(p-2))/((p-3)^2+(p-2)^2)$
as $q$ goes to infinity at fixed $p$.  To see this, we look at how the
growth rate $\lambda_{\tau C(p,q)}$ grows in this limit.  We find it grows to infinity as $(p-2)q$ so that again the
factor $\lambda_{\tau C(p,q)}/(\lambda_{\tau C(p,q)}-1)$ in the code rate goes to one.  Similarly, the growth vector becomes parallel to the edge vector
$\vec{u}_{\tau C(p,q)}=q\vec{e}_{\tau C(p,q)}$, up to corrections of order $1/q$.  Thus the number
of tiles meeting around a vertex $q$ cancels from the ratio
($\vec{u}_{\tau C(p,q)}\cdot\vec{t})/(\vec{u}_{\tau C(p,q)}\cdot\vec{e}_{\tau C(p,q)})$ giving the result that the
code rate falls to the non-zero, $q$ independent function
$(\vec{e}_{\tau C(p,q)}\cdot\vec{t})/(\vec{e}_{\tau C(p,q)}\cdot\vec{e}_{\tau C(p,q)})$ in the limit that $q$ goes to infinity at fixed $p$.  Thus, we have shown that the holographic code grown with tile completion on the $\{p,q\}$-tiling of the hyperbolic plane by regular $p$-gons meeting $q$ around a vertex has code rate $\chi_{\tau C(p,q)}$ which falls as $p$ increases at fixed $q$ and as $q$ increases at fixed $p$, for all $p$ and $q$ such that $1/p+1/q<1/2$.

\section{Conclusion}
Holographic codes living on the tiles of the $\{p,q\}$ tessellations of
the hyperbolic plane with $p$-sided regular tiles meeting $q$ around a
vertex have code rate $\chi_{\tau(p,q)}\leq \ell_{p,q}/a_{p,q}$ for a
code grown layer by layer using inflation rule $\tau(p,q)$.  Here the
length of the sides of the tiles $\ell_{p,q}$ and their area $a_{p,q}$
combine to give an upper bound on the code rate from hyperbolic
geometry.

We find the tiling with the best code rate bound for holographic codes on hyperbolic tessellations with regular $p$-sided tiles for all $p\ge 3$.  The best bound falls quickly with $p$ as does the optimal number $q_{opt}(p)$ of
tiles meeting around each vertex.  We show that there exists the range $(q_{min}(p),q_{max}(p))$ within which {\em all} holographic codes grown on regular $\{p,q\}$-tilings of the hyperbolic plane perform quantum error correction.  In particular we find $q_{min}(p)=1+\lfloor2+4/(p-2)\rfloor$ includes all tilings allowed by hyperbolic geometry for all $p>3$ and we show that $q_{max}(p)$ grows $O(\exp(\pi p/2))$ exponentially with $p$.  
  
Finally, the code rate of the holographic codes grown with tile
completion on the $\{p,q\}$-tessellation of the hyperbolic plane was
computed.  The triangle codes have code rate greater than one and cannot perform quantum error correction while those with tiles having more than three sides have code rate less than one and can perform quantum error correction.  The computed code rates obey the upper bound from hyperbolic geometry.
\begin{acknowledgments}
The authors gratefully acknowledge helpful discussions with Latham Boyle, Alan Fuchs, Ray Aschheim, and Fang Fang.  Funding was provided by Quantum Gravity Research (Research Proposal: ``Quasi-Crystalline Tensor Networks").    
\end{acknowledgments}

\bibliography{coderate}

\end{document}